\documentclass{llncs}

\usepackage{wine} 
\usepackage[colorinlistoftodos]{todonotes} 

\begin{document}

\title{Annotation based automatic action processing}
\author{Elias K\"arle \and Dieter Fensel}
\authorrunning{Elias K\"arle et al.} 
%
\tocauthor{Elias K\"arle and Dieter Fensel}
\institute{Semantic Technology Institute, Universit\"at Innsbruck, Innsbruck, Austria,\\
\email{\{elias.kaerle, dieter.fensel\}@sti2.at}
}

\maketitle              

\begin{abstract} 
With a strong motivational background in search engine optimization the amount of structured data on the web is growing rapidly. The main search engine providers are promising great increase in visibility through annotation of the web page's content with the vocabulary of schema.org and thus providing it as structured data. But besides the usage by search engines the data can be used in various other ways, for example for automatic processing of annotated web services or actions. In this work we present an approach to consume and process schema.org annotated data on the web and give an idea how a best practice can look like.
\keywords{semantic web, semantic web services, open data, schema.org}
\end{abstract}

\section{Introduction}
\label{sec:introduction}
The introduction of intelligent personal assistants (IPAs), like Amazon's Echo, Apple's Siri, Google's Allo or Microsoft's Cortana, is about to fundamentally change the way we search for information or consume content on the web. With the introduction of schema.org in 2011 the web received a de-facto standard for structuring content on the web and make it machine read- and interpretable. Annotating a website with schema.org is currently a common SEO\footnote{Search Engine Optimization} practice to increase visibility. 
Yet the semantically enriched data can also be used by third party software and hence make the website a database-like knowledge source.

With this work in progress paper we present the idea of using schema.org annotated data on the web to automatically process information and to execute schema.org actions in the manner of semantic web services. Based on a predefined set of websites (URLs of websites of similar content) our system collects schema.org annotated data, stores it and provides the data over a likewise semantically annotated API to third party software, like before mentioned IPAs, chatbots or alike.

In four steps we are (1) collecting the data of a predefined set of websites and (2) storing it into a document store or graph database. In step (3) we define a layer for data retrieval and step (4) defines the API for third party consumption and describes the API with schema.org actions. As a showcase we are using an example from the tourism sector. Tourism is a convenient example because in an analysis we found out, that this sector is increasingly adapting schema.org for hotels \cite{karle2016there}, events, POIs and alike, and because we know of the existence of an almost fully annotated destination marketing organization platform \cite{akbar2017massive}.

\section{Related Work}
Some of the steps mentioned in Section \ref{sec:introduction} are already well established and discussed in the literature, like crawling and data storage. Others are not sufficiently covered, like data retrieval and reconciliation, and will be the matter of contribution in this paper. This sections presents literature about both areas. In \cite{harth2007yars2} the authors describe YARS2, a system to query content in a structured data graph built upon information from various websites. In \cite{ding2004swoogle} the authors present Swoogle a, as the name suggests, Google for the semantic web (before Google itself started operating in the field of the semantic web). As opposed to the two ideas we do not want to provide a holistic semantic web search engine but an endpoint for domain specific annotated data and annotated web services from a predefined set of URLs which we then provide not over a user interface but over an API. Very early attempts of extracting structured and semi-structured data from the web can be found in \cite{abiteboul1997querying} where the authors were creating wrappers based on templates to extract content from the web. Our approach is targeting fully structured data in RDFa, Microdata or JSON-LD format. The data found by the crawler will be translated to JSON-LD and, due to its JSON nature, stored in the NoSQL database MongoDB. The advantages of using NoSQL over classical RDBMSs are described in \cite{moniruzzaman2013nosql}. A comparison of MongoDB to Cassandra, another widely distributed NoSQL database can be found in \cite{abramova2013nosql} and justifies the advantages of the use of MongoDB for our use case. Another way to store semantic data would be, due to its triple nature, a graph database as described in \cite{angles2005querying}, which we consider as a storage for the next step of this paper. The challenges of information retrieval, especially in large graph databases, are described in \cite{weikum2009database}, a solution for semantic data reconciliation can be found here \cite{gal2005framework}. Even though our approach is not using a graph database but the above mentioned document database (MongoDB) similar challenges occur and a lot can be learned from the mentioned works. The identification of (web-) services will use techniques of semantic web service discovery as in \cite{klusch2006automated}. The publication of the structured data found on the web will be enabled by semantically annotated web services, as opposed to the ideas found in \cite{fensel2006enabling} and \cite{ankolekar2002daml} we are using schema.org actions as a light weight semantic web service annotation language.

\section{Methodology}
This section presents the methodology of the four steps mentioned in the introduction in more detail. The general idea is to collect the structured data which is available on certain websites and process the web services annotated with schema.org in an automatic way. Depending on the domain those web services can be a booking of a touristic service, like a hotel room or a ski course, a purchase of a product in a web shop or others. A precondition for working with schema.org annotated data is an efficient and automated way to publish it for website owners. We are also working on means for automatic annotation publication (like described in \cite{karle2017semantify} amongst others) but that would exceed the scope of that paper. The starting point for the process is a predefined collection of URLs of websites containing schema.org annotated data. First we collect that structured data and identify the type of data and the presence of web services by web service discovery (see therefore \cite{klusch2006automated}). The collecting frequency is hereby dependent on the sort of data (static, dynamic or active data). To further process the data and to perform reasoning we store everything we found in a database. In the first step we are using the NoSQL database MongoDB because it stores data in form of JSON which is the exact same format most of the annotations of our use case are in. Hereby we have to take care of avoiding redundancies, combine newly found entities with already existing entities in the graph by entity resolution and make meaningful subgraphs. By the time our store is populated, we can query the data from the document store. The challenge here is to find exactly the data the user is looking for, even though the data is heterogeneous and might be incomplete and erroneous. The methods of information retrieval, semantic reconciliation and heuristic classification have promising approaches and will be used here. To make the data accessible for third party software it will be published through an application programming interface (API). This interface will be described as a semantic web service itself with schema.org actions, and take search parameters as an input and return schema.org objects as results which can be used for further interaction with the API.

\section{Use Case}
As a show case we take a domain with a high density of many well and fully annotated websites, the tourism sector and the destination marketing organization (DMO) of Mayrhofen (annotated during the work on \cite{akbar2017massive}). We collect the structured data on the DMO's website, store it in MongoDB and make an Alexa skill which is connected to our web service. So when a user asks Alexa about an available hotel room, the request will be forwarded to our API. Our API then queries our database and returns an answer, containing a schema.org object with room offers. The offer object contains further references to actions (API requests), for example a reservation request for the room or a booking action. Those requests are then referred to the source where our system initially found them, for example the accommodation provider's internet booking engine. So our system acts like a proxy and the actual business logic stays with the accommodation provider.

\section{Conclusion \& Ongoing Work}
This work-in-progress paper describes a way to collect schema.org annotated data from websites, store the data in a document- or graph database and provide the information over an API for automatic consumption of content and execution of actions. The ongoing work covers all four steps of the methodology to raise the development level of this research idea to a working prototype for real life application.

\bibliographystyle{splncs}
\bibliography{bib}

\end{document}